\def\etbr{$\kappa$-(BE\-DT\--TTF)$_2$\-Cu\-[N\-(CN)$_{2}$]Br}
\def\etcl{$\kappa$-(BE\-DT\--TTF)$_2$\-Cu\-[N\-(CN)$_{2}$]Cl}
\def\etbrcl{$\kappa$-(BE\-DT\--TTF)$_2$\-Cu\-[N\-(CN)$_{2}$]\-Br$_{x}$Cl$_{1-x}$}
\def\cm{cm$^{-1}$}
\documentclass[prb,twocolumn,showpacs]{revtex4}
\usepackage[dvips]{graphicx}
\begin{document} 
\title{Bandwidth-controlled Mott transition in $\kappa$-(BEDT-TTF)$_2$Cu[N(CN)$_{2}$]\-Br$_{x}$Cl$_{1-x}$:\\
Optical studies of correlated carriers}
\author{Michael Dumm}
\author{Daniel Faltermeier}
\author{Natalia Drichko}
\author{Martin Dressel}
\affiliation{1.~Physikalisches Institut, Universit\"{a}t
Stuttgart, Pfaffenwaldring 57, D-70550 Stuttgart Germany}
\author{C\'ecile M\'eziere}
\author{Patrick Batail}
\affiliation{ Laboratoire CIMMA, UMR 6200 CNRS-Universit\'{e} d'Angers,
B\^{a}t. K, UFR Sciences, 2 Boulevard Lavoisier, F-49045 Angers, France}
\date{\today}
\begin{abstract}
In the two-dimensional organic charge-transfer salts \etbrcl\ a
systematic variation of the Br content from $x=0$ to 0.9 allows us
to tune the Mott transition by increasing the bandwidth. At
temperatures below 50~K, an energy gap develops in the Cl-rich
samples and grows to approximately 1000~cm$^{-1}$ for $T\rightarrow
0$. With increasing Br concentration spectral weight shifts into the
gap region and eventually fills it up completely. As the samples
with $x=0.73, 0.85$ and 0.9 become metallic at low temperatures, a
Drude-like response develops due to the coherent quasiparticles.
Here, the quasiparticle scattering rate shows a $\omega^2$
dependence and the effective mass of the carriers is enhanced in
agreement with the predictions for a Fermi liquid. These typical
signatures of strong electron-electron interactions are more
pronounced for compositions close to the critical value $x_c\approx
0.7$ where the metal-to-insulator transition occurs.
\end{abstract}

\pacs{
71.30.+h, 
71.10.Hf, 
74.70.Kn,   
74.25.Gz 
}

\maketitle
%
%

\section{Introduction}
One of the most intriguing issues in condensed-matter physics is the
transition from a metal to an insulator driven by electronic
correlations. Why does an electron in a crystal change from
itinerant to localized behavior when a control parameter such as
magnetic field, doping or pressure is varied? For a system with a
\begin{figure}
\centering
\includegraphics[width=80mm]{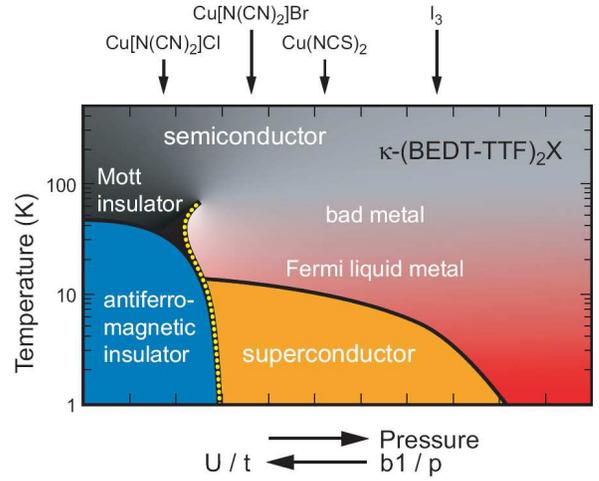}
\caption{\label{fig:ETphasediagramII}Schematic phase diagram of $\kappa$-(BEDT-TTF)$_2X$. The on-site Coulomb repulsion with respect
to the hopping integral $U/t$ can be tuned either by external pressure or modifying the
anions $X$. The arrows indicate the approximate position of $\kappa$-phase salts with $X$ = Cu\-[N\-(CN)$_{2}$]Cl, Cu\-[N\-(CN)$_{2}$]Br, Cu(NCS)$_2$, and I$_3$
at ambient pressure, respectively. The bandwidth controlled phase transition between the insulator and the
Fermi liquid/superconductor can be explored by gradually replacing Cl by Br in
\etbrcl. Here $c$ and $a$ are the lattice parameters; $b1$ and $p$ indicate intra- and interdimer transfer integrals, respectively.
}
\end{figure}
half-filled conductance band this problem is known as the Mott
transition, one of the central problems of strongly-correlated
electrons.\cite{Mott74} While studies of the  influence of
electron-electron interactions in materials with open $d$ and $f$
electron shells have a long history,\cite{Fulde93,Grewe91,Imada98}
only recently it was realized that also in molecular conductors
(where the charges originate from molecular orbitals) electronic
correlations are very
significant.\cite{Jerome94,Ishiguro98,McKenzie98,Seo04,Dressel04,Fukuyama06}
In transition-metal oxides, the Coulomb interaction is crucial for
any understanding of the unconventional metallic and superconducting
properties as well as the vicinity to magnetic order; this is most
pronounced in the underdoped cuprates.\cite{Imada98} However, in
many regards organic conductors turn out to be superior model
systems to study certain effects of electron-electron interaction
since their properties can be more easily tuned by (physical or
chemical) pressure. Varying $U/t$, where $U$ is on-site repulsion
and $t$ is a  transfer integral, opens the road to the
bandwidth-controlled Mott transition.

The family of half-filled organic conductors
$\kappa$-(BEDT-TTF)$_2X$ (anions $X=$~Cu(CN)$_3$, Cu[N(CN)$_2$Cl,
Cu[N(CN)$_2$Br, Cu(SCN)$_2$, and I$_3$)\cite{remark0} has a
particulary  rich phase diagram as a function of pressure and
temperature as depicted in Fig.~\ref{fig:ETphasediagramII}. The
abscissa of this phase diagram can be interpreted as the  variation
of the relative Coulomb interaction $U/t$. Similar to external
pressure, a variation of anions also changes the bandwidth and thus
reduces $U/t$. For large values of $U/t$ the half-filled system
becomes a Mott insulator. This behavior is observed in \etcl: at
ambient conditions the narrow-gap semiconductor gradually gets
insulating when cooled below 50~K. The application of external
pressure shifts the compound across the phase boundary. It becomes
metallic and even undergoes a superconducting transition, very
similar to the Br analog at ambient pressure.  Recently, the
critical behavior in the vicinity of the metal-insulator transition
and the critical endpoint was thoroughly investigated by dc
measurements  under external pressure and magnetic
field.\cite{Lefebvre00,LIM03,KAG04} In the present study we
gradually substitute Cl by the isovalent Br in the anion layers and
obtain the series \etbrcl\ crossing over from the Mott insulator
with antiferromagnetic ground state (Cl-compound) to a Fermi liquid
which becomes superconducting at $T_c=12$~K (Br-compound).

Over the last decade dynamical mean-field theory has been
established as a powerful tool to study the physical properties of
highly-correlated electron systems.\cite{Georges96,Kotliar04} To
understand the properties of $\kappa$-phase BEDT-TTF-based salts,
Kino and Fukuyama\cite{Kino96} suggested to model them by a
triangular lattice of BEDT-TTF dimers with one hole per site and
hopping between the dimer sites $t_1$ and $t_2$ and on-site
repulsion $U$. A Mott-type metal-insulator transition occurs at some
critical value of the relative Coulomb repulsion $(U/t)_c$. Merino
and McKenzie\cite{Merino00,Merino08} evaluated the transport
properties of the metallic side of the phase diagram of these
half-filled systems using a dynamical mean-field treatment (DMFT) of
the Hubbard model with strong on-site Coulomb repulsion $U \approx
W$ (with $W=10t$ being the bandwidth for the frustrated square
lattice in tight-binding approximation). They show that the optical
conductivity exhibits a zero-frequency mode (Drude peak) at low
temperatures, while it is suppressed above some coherence
temperature $T_{\rm coh}$, meaning close to a Mott metal-insulator
transition the quasiparticles are destroyed for $T>T_{\rm coh}$.

Optical investigations give a respective experimental insight into
the dynamics of the conduction electrons, including the existence of
the coherent and incoherent quasiparticle
response.\cite{Eldridge91b,Kornelsen92,Haas00,Dressel03,Basov05,Drichko06,Sasaki08}
In Ref.~\onlinecite{Faltermeier07} we presented the  reflectivity
data (frequency range: 50 to 10\,000~\cm, temperature range: 5 to
300~K) received on single crystals of \etbrcl\ ($x= 0$, 0.4, 0.73,
0.85, and 0.9). The interpretation of the frequency-dependent
conductivity $\sigma_1(\omega)$ which is shown in the left panels of
Fig.~\ref{fig:Spek} for $E\parallel c$ was performed in terms of two
contributions: charge transfer inside the dimer ``lattice sites''
and interdimer charge transfer by correlated charge carriers. At
ambient temperature the frequency-dependent conductivity
$\sigma_1(\omega)$ is dominated by a broad absorption peak located
at frequencies around 2000 - 3500~\cm. Down to $T=50$~K no
Drude-like contribution to the optical conductivity is present
although the crystals are moderately good conductors. At even lower
temperatures a zero-frequency contribution is observed for \etbrcl
with $x$ = 0.73, 0.85, and 0.9, but not for the lower concentrations
of Br where an energy gap opens at the metal-insulator transition.
Using the cluster model suggested by Rice, Yartsev and
coworkers,\cite{Rice76,Yartsev93,YAR96} we showed that the
electronic band with a maximum at around 3500~\cm\ ($E\parallel c$)
and the narrow features at frequencies of BEDT-TTF molecular
vibrations are due to the charge transfer within a dimer coupled to
$A_g$ vibrations of BEDT-TTF. This analysis permits to disentangle
the intradimer and interdimer carriers contributions to the spectra,
as demonstrated Refs. \onlinecite{Faltermeier07} and
\onlinecite{Dressel09} whereas the latter account for the
Mott-Hubbard physics due to strong electronic correlations.

In the present study we closer inspect the transition into the
Mott-insulating state in \etbrcl\ driven either upon lowering the
temperature or decrease of chemical pressure. We confine ourselves
to a detailed analysis of one of the in-plane polarizations (the
c-axis) where it is easier to disentangle the different
contributions to the spectra. However, an analysis of the $a$-axis
spectra leads to qualitatively comparable results; the response of
the strongly-correlated carriers is basically isotropic in the
conducting plane.\cite{Faltermeier07} Starting with the proposed
density of states, we will discuss the following aspects of the
Mott-Hubbard model system: the appearance of an energy gap and a
quasiparticle peak at low temperatures and their evolution as
function of temperature and effective Coulomb interaction, the
suppression of the spectral weight across the transition, and the
Fermi-liquid response of the metallic state in the vicinity of the
Mott transition.

\section{Results and Discussion}

\subsection{Analysis of the complete spectra}

\subsubsection{Redistribution of spectral weight}
\label{sec:spectralweight}
\begin{figure}
\includegraphics[width=85mm]{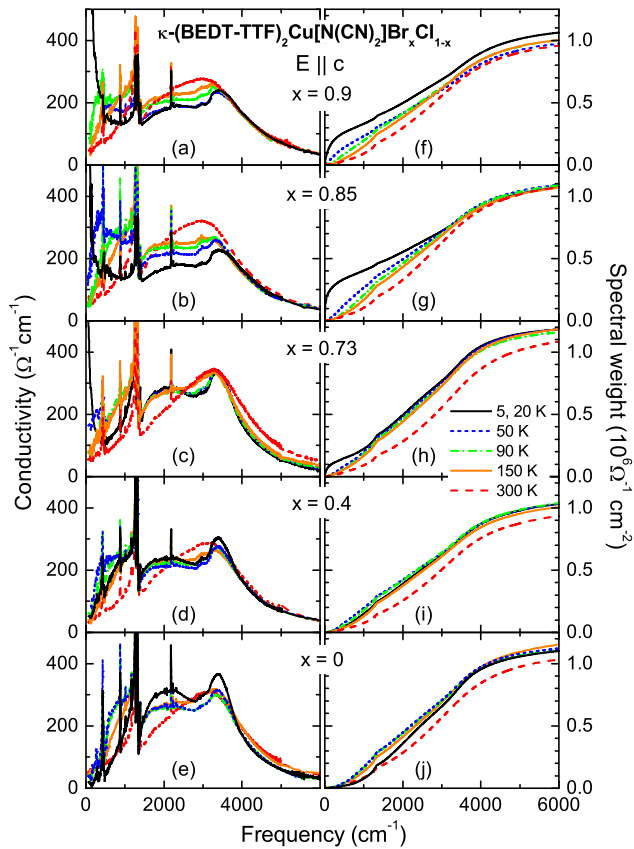}
\caption{\label{fig:Spek}(Color online)
Left frames: experimental spectra of the real part of the complex conductivity $\sigma_1$ of \etbrcl\ for
the in-plane $E\parallel c$ measured at different temperatures;
right frames: spectral  weight as a function of cut-off frequency $\omega_c/(2\pi c)$ calculated according to Eq.~(\protect\ref{fsum}).
From the top to the bottom panels, (a) - (e) and (f) - (j) respectively,
the Br content is reduced from 0.9 to 0.}
\end{figure}
Figs.~\ref{fig:Spek}(a)-(e) show the optical conductivity for $E
\parallel c$ in the measured frequency range. Within this range we
observe all the studied processes, the intradimer transitions and
the interdimer transitions which show up as Hubbard bands and
Drude-like peak, depending on the compound and temperature. The
relatively distinct drop in reflectivity around 4000 to 5000~\cm\
shown in Ref.~\onlinecite{Faltermeier07} infers that the optical
conductivity up to these frequencies is governed by the electrons in
the conduction band formed by the overlap of BEDT-TTF orbitals in
the layer. The one-dimensional tight-binding model illustrates the
proportionality between the value of the spectral weight and
transfer integral $t$: $\int_0^{\infty}\sigma_1(\omega)\,{\rm
d}\omega=
\frac{2td^2e^2}{\hbar^2V_m}\sin\left\{\frac{\pi}{2}\rho\right\}$,
where $d$ is the inter-molecular distance, $V_m$ denotes the volume
per molecule, and the electrons per site are given by $\rho$. Due to
the triangular lattice, there is no single transfer integral that
can define the width of the band; nevertheless the equation still
gives the correct idea, that up to about 6000 \cm\ we are dealing
with intra-band transitions.\cite{remark1}

The frequency-dependent spectral weight is given by the integral
\cite{DresselGruner02}
\begin{equation}
I_\sigma(\omega_c)=
\int_0^{\omega_c} \sigma_1(\omega)\, {\rm d}\omega
= \frac{\pi e^2}{2 m^{*}_{\rm sum}}n(\omega_c)
\quad ,
\label{fsum}
\end{equation}
where $\omega_c$ is a cut-off frequency and $m^{*}_{\rm sum}$ an
effective mass which is equal to the optical band mass $m_{\rm
b,opt}$ in the non-interacting case; $n(\omega_c)$ indicates the
density of carriers contributing to the conductivity up to
$\omega_c$. First of all, within the experimental uncertainty the
spectral weight is approximately the same for all temperatures and
Br concentrations when going up to $\omega_c/(2\pi c)=6000$~\cm\ or
higher,\cite{remark2} as demonstrated in Fig.~\ref{fig:Spek}(f)-(j).
The conservation of the spectral weight within the band suggests
that we can ignore the higher-frequency interband transitions and
focus our attention on the spectral range below 6000~\cm. Within
this region, the frequency redistribution of the spectral weight
with varying temperature and Br concentration shows from which
energy range what kind of charge carriers contribute to the optical
response [Eq.~(\ref{fsum})]. In conventional metals most of the
optical weight is concentrated in the Drude peak and
$I_\sigma(\omega_c)$ should quickly saturate with frequency, which
is not the case in the studied compounds.

The steps around 500 and 1200~\cm\  in all the $I_\sigma(\omega_c)$
curves are due to the strong emv-coupled molecular vibrations. Most
important, the distribution of the spectral weight significantly
changes for the different samples. With lowering the temperature, a
shift to lower frequencies occurs that is much more pronounced in
the crystals with high Br content. In the salts with $x =0.85$ and
0.9, a strong increase of the spectral weight is observed below
$\omega_c/(2\pi c)=1000$~\cm\ and $T=150$~K, while in
$\kappa$-(BEDT-TTF)$_2$\-Cu\-[N\-(CN)$_{2}$]\-Br$_{0.73}$Cl$_{0.27}$
this shift is less prominent.  Contrary to this tendency, below
150~K the far- and mid-infrared spectral weight decreases in the
samples with $x = 0.4$ and 0. It should be pointed out that the
difference between the temperature behavior of the samples is
observed only below 3000~\cm, i.e.\ in the spectral region of the
correlated-carriers contribution, which will be analyzed below.
\begin{figure}[b]
\includegraphics[width=80mm]{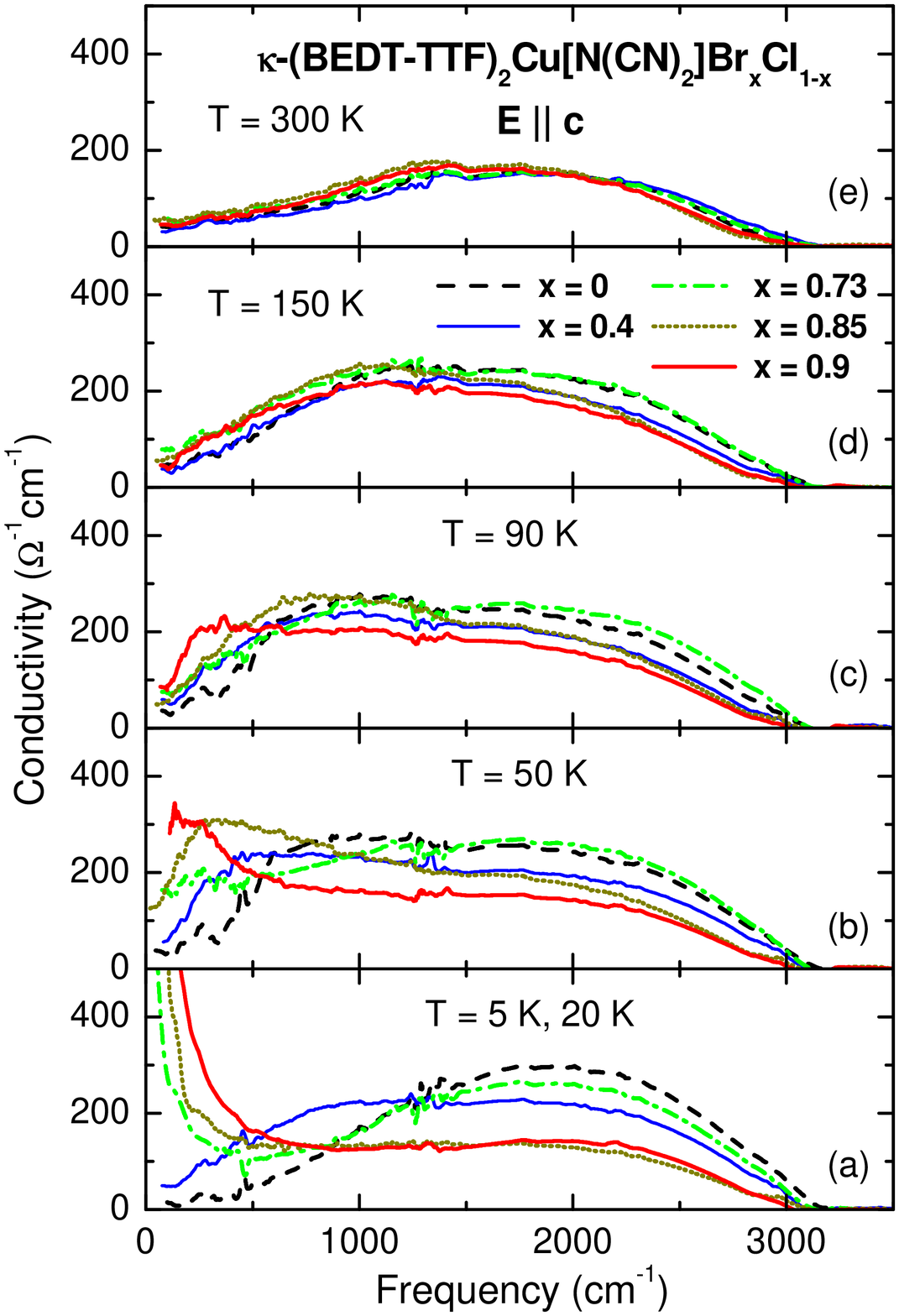}
\caption{(Color online) Frequency dependence of the
 conductivity $\sigma_1$ ($E\parallel c$) due to correlated charge carriers after the
contributions from intradimer transitions and vibrational modes are subtracted. The
frames (a) - (e) show spectra of \etbrcl\ with different Br
concentrations $x$ at various temperatures as indicated.\label{fig:condsubtracted}}
\end{figure}

\subsubsection{Mott-Hubbard system}
\label{sec:MottHubbardsystem} 
\begin{figure}[h]
\rotatebox{270}{\scalebox{0.55}{\includegraphics{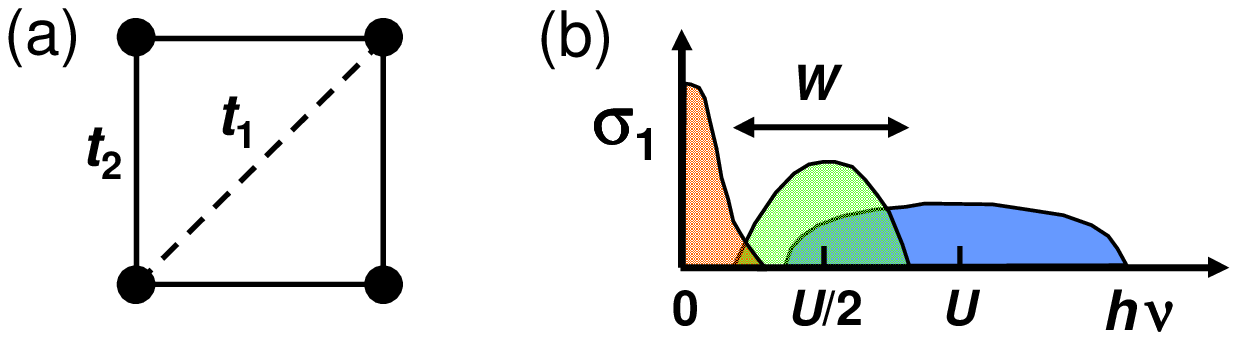}}}
\caption{(a) Anisotropic triangular lattice used in theoretical
calculations of the electronic properties; the ratio $t_1/t_2=0.8$.
(b) A schematic view of conductivity spectrum predicted by DMFT for a metallic compound close to the Mott transition.}\label{fig:lattice}
\end{figure}

For the inspection of the dynamics of
the correlated charge carriers, the features due to the intradimer
transitions were extracted from the experimental spectra of both
$\sigma_1$ and $\sigma_2$ in a Kramers-Kronig consistent way,
following the ideas and procedure developed and discussed in
Refs.\onlinecite{Faltermeier07} and \onlinecite{Dressel09}; the
resulting $\sigma_1$ spectra are presented in
Fig.~\ref{fig:condsubtracted}. At lowest temperatures shown in panel
(a) of Fig.~\ref{fig:condsubtracted}, we can distinguish two main
contributions: (i) a finite-frequency part (centered around
2000~\cm), which is the sole contribution in the case of low Br
content $x=0$ and 0.4, and which becomes weaker as $x$ increases;
(ii) a zero-frequency peak for the large-$x$ samples due to coherent
particle response. These two basic features are in good agreement
with a theoretical prediction for a  Mott-Hubbard system
\cite{Rozenberg95} as sketched in Fig.~\ref{fig:lattice}(b): a
Drude-like peak at $\omega=0$, and a broad absorption band at about
$U$; while we do not observe a mid-infrared band at $U/2$ in our
spectra. Assuming a frustrated square lattice as depicted in
Fig.~\ref{fig:lattice}(a) with hopping parameters $t_1=0.024$~eV and
$t_2=0.03$~eV, and $U=10|t_2|=0.3$~eV, Merino and
McKenzie\cite{Merino00,Merino08} could quantitatively describe the
experimental findings. The broad band at 2000~\cm\ is attributed to
electronic transitions between the two Hubbard bands. Interestingly,
the position of the Hubbard transition does not change with
temperature and upon Br substitution. This implies that it does not
depend on the transfer integral $t_2$ and the degree of frustration
$t_1/t_2$, but only on the Coulomb interaction $U$ that is identical
for all materials.

If the temperature  increases [panels (b) to (c) of
Fig.~\ref{fig:condsubtracted}], the coherent carriers peak
disappears for the high Br concentrations, showing that no coherent
transport is possible anymore. This results in the so-called
``bad-metal'' behavior, where the dc resistivity still increases
with temperature up to the maximum at about 100~K but no band-like
transport occurs. An evaluation of the Hubbard model by dynamical
mean-field theory reproduces the signatures of the gradual
destruction of quasiparticles as the temperature passes $T_{\rm
coh}$. At 150~K and above [panels (d) and (e) of
Fig.~\ref{fig:condsubtracted}], independently of the Br content a
semiconducting behavior is recovered which is characterized by a
negative slope of $\rho$ versus $T$ (Fig.~3 of
Ref.~\onlinecite{Faltermeier07}).

While these main features of the metallic samples are in good
agreement with theory, the overall picture of the conductivity
spectra of the correlated carriers as function of temperature and Br
content $x$ adds new information. At high temperatures, $\sigma_1
(\omega)$ does not depend much on Br content $x$, and a shift of
spectral weight to lower frequencies is observed for {\em all} the
compounds when going from 300~K to 150~K. The behavior becomes
qualitatively different at 50~K as the higher-frequency contribution
gets noticeably reduced and a peak emerges below 500~\cm\ for $x=$
0.9, 0.85, and 0.73. At lowest temperature, 20~K and 5~K, this peak
evolves into the coherent carriers peak; the finite-frequency band
becomes so reduced that it can be barely distinguished from the
higher-frequency wing of the Drude contribution. For lower Br
concentration, the intensity of the transitions between the Hubbard
bands at about 2000~\cm\ remains the same, though the contribution
at about 300~\cm\ rises as temperature increases above 35~K and the
systems cross over into the semiconducting state.

\begin{figure}
\includegraphics[width=55mm]{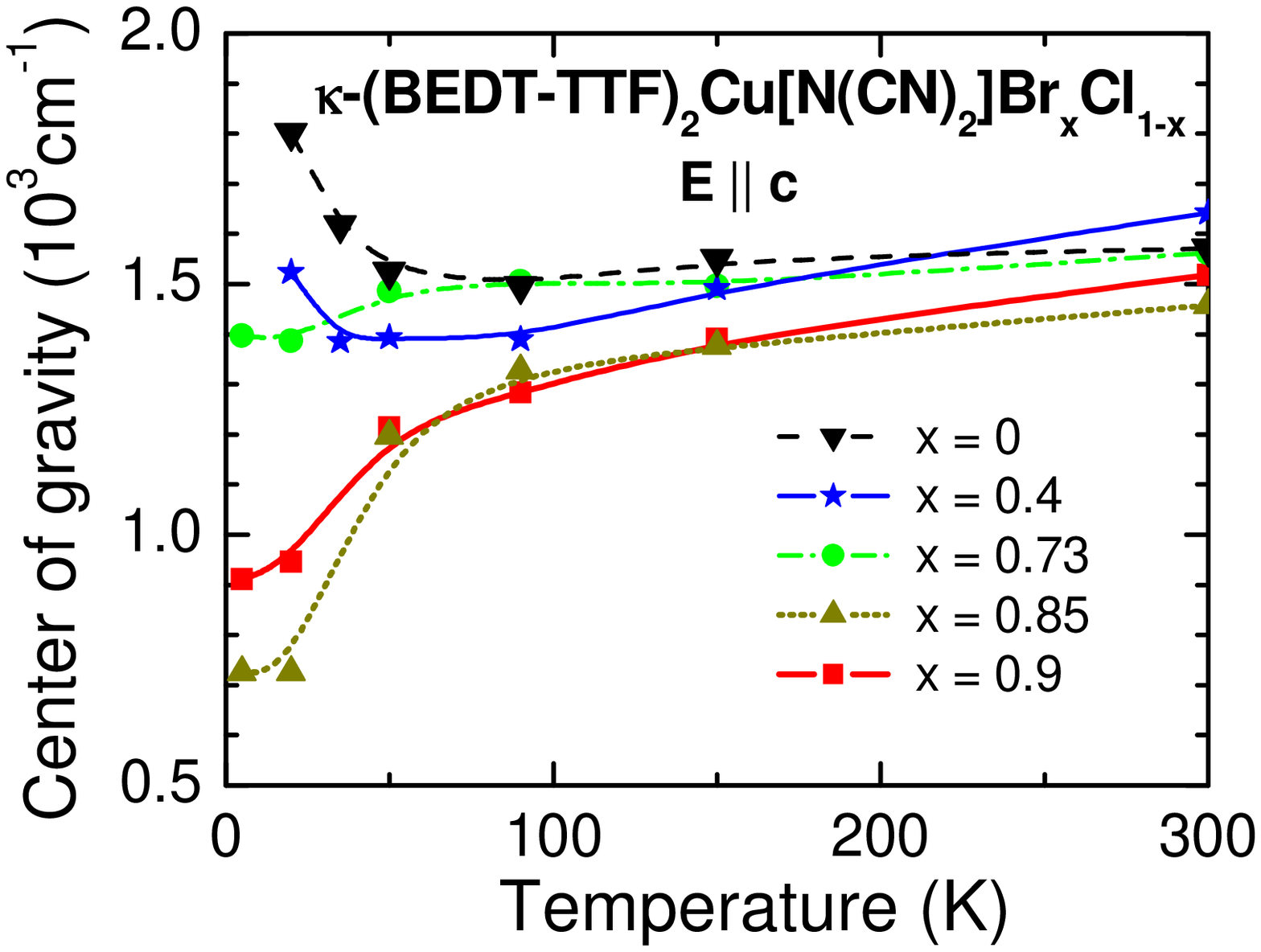}
\caption{\label{fig:center}(Color online) Shift of the center of gravity in the conductivity spectra of the correlated charge carriers with
temperature for different $x$ in \etbrcl\ for $E\parallel c$. The lines correspond to spline fits.}
\end{figure}

The spectral-weight shift towards lower frequencies on cooling from
300 to 50~K can be interpreted as a signature of ``getting closer''
to a metallic state; this is illustrated by the monotonic decrease
of the center of gravity of the spectra shown in
Fig.~\ref{fig:center} as a function of temperature, where the values
are calculated from Fig.~\ref{fig:condsubtracted}. At lower
temperatures the behavior is distinctively different: the shift
becomes more pronounced for the metallic samples with $x = 0.85$ and
0.9, while the opening of the Mott-Hubbard gap causes an upward
shift of the center of gravity for $x=0$ and 0.4. Interestingly, the
sample with $x = 0.73$ is positioned between both limiting cases,
but at the lowest temperatures the coherent particle response
appears.

\subsection{Density of states}
\begin{figure}
\rotatebox{270}{\includegraphics[width=58mm]{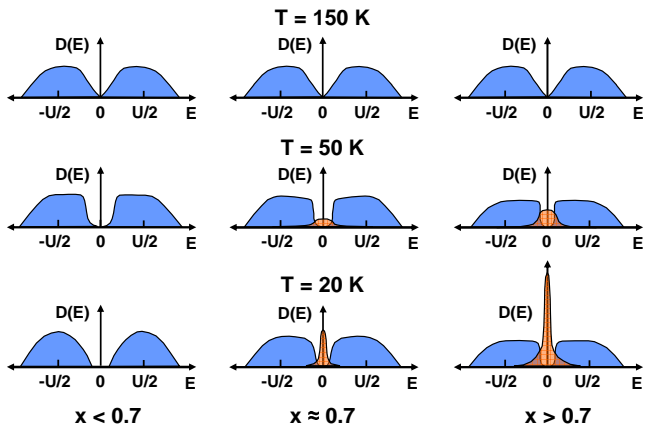}}
\caption{\label{fig:dos}(Color online) Density of states at $T = 150$~K,
50~K, and 20~K as proposed for the low Br-doped compounds ($x < 0.7$),
the crystal with $x \approx 0.7$, and the highly Br-doped samples ($x < 0.7$) from the optical conductivity.
The zero-frequency mode is shown in orange color and the finite-frequency modes in blue color.
The temperatures represent the metallic or Mott-insulating region (20K), the narrow-gap semiconducting high-temperature region (150K)
and the crossover regime between the former which is often referred to as ``bad-metal'' or ``bad-semiconductor'' region. }
\end{figure}

For further understanding and comparison to the theory, we sketch in
Fig.~\ref{fig:dos} the density of states (DOS) as proposed from the
above discussion of our data. We chose three typical temperature
(low temperature, around 50 K, and high temperature) and correlation
regions (high Br content, $x = 0.73$, and low Br content) in order
to cover all relevant parts of the phase diagram
(Fig.~\ref{fig:ETphasediagramII}). The upper and lower Hubbard bands
and, accordingly, the optical transitions between them are present
for all the temperatures and correlation values.
At 150 K and higher temperatures these bands are more or less
symmetric and seem to extend down to zero frequency. When
temperature is  decreased down to 90 and 50 K, respectively, the
center of gravity of the finite-frequency bands moves gradually to
lower frequencies (Fig.~\ref{fig:center}). From the optical
conductivity (Fig.~\ref{fig:condsubtracted}) it is obvious that the
spectral weight shifts from the mid-infrared into the far-infrared
region, however, the appearance of the strong absorption edge below
500~\cm\ signals a much more abrupt onset of the absorption band at
these temperatures. The onset frequency is smaller for the compounds
with lower correlation values (higher Br dopings) where the sharp
edge of the finite-frequency absorptions and the emerging
quasiparticle peak add up to a maximum in the total DOS just above
the gap; such kind of behavior is predicted by cluster DMFT
calculations.\cite{Kyung06,Park08,Ohashi08} At lowest temperatures,
the coherent carriers peak dominates the DOS close to the Fermi
energy in the salts with $x > 0.7$ which corresponds to the
Drude-like response in the optical conductivity.

The calculations in Refs.~\onlinecite{Kyung06,Park08,Ohashi08} were
performed for low temperatures in the Mott-insulating phase close to
the Mott transition in a half-filled two-dimensional system. They
suggested that besides the Hubbard bands situated at $|E| = U/2$,
there might be more contributions to the broad bands at $|E| > 0$. A
particular attention was given to short-range correlations which
cause an additional band in the spectra. In contrast to single-site
calculations, cluster DMFT reveals these short-range correlations
which reduce the critical Coulomb repulsion $U$; most important in
this context, additional low-frequency excitations are expected due
to local singlet formation.

There remains a considerable discrepancy  between our experimental
data and single-site DMFT calculations: the absence of the $U/2$
peak related to excitations from the Hubbard band to the
quasiparticle peak [compare Figs.\ref{fig:condsubtracted}(a) and
\ref{fig:lattice}(b)]. Although there is an overlap between the
higher-frequency part of the Drude peak and the broad band of
transitions between Hubbard bands in the experimental data which
might hide a weak absorption feature, it is obvious that this
contribution is not as strong as predicted by theory. However, in
this regard it is important to mention that optical experiments do
not resolve the $k$-dependence which causes a smearing of the
spectral features. Additionally, the single-band approach by the
theory might be a reason for the deviations.

\subsection{Metal-insulator transition}

\subsubsection{Energy gap\label{sec:gap}}
A characteristic signature of the metal-insulator transition is the
opening of an energy gap in the excitation spectrum. Our experiments
on crystals with different Br-concentrations provided the
opportunity to probe the transition into the Mott-insulating state
as function of temperature and correlation: at low Br content, a
transition from a narrow-gap semiconducting\cite{remark7} into a
Mott-insulating state occurs on lowering the temperature, while at a
fixed low temperature the system crosses over from a metallic into a
Mott-insulating state on the increase of the relative Coulomb
repulsion $U/t$.
\begin{figure}[h]
\includegraphics[width=85mm]{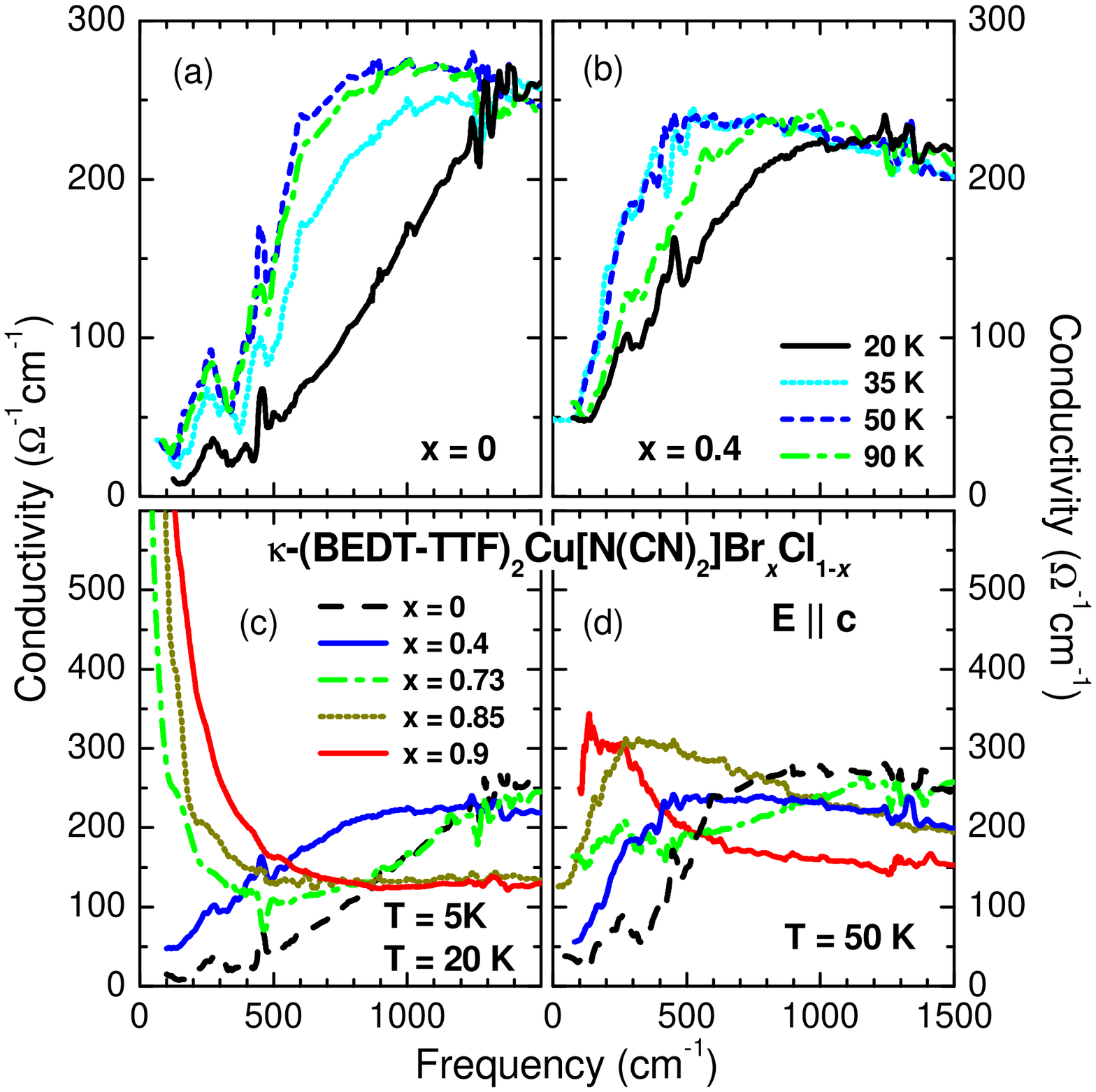}
\caption{\label{fig:gapa}(Color online)  Comparison
of the development of the optical gap with changing the Br
concentration and temperature for $E\parallel c$. The panels show the frequency-dependent conductivity $\sigma_1$
after subtracting contributions due to localized charge carriers and vibrational modes.
(a) and (b): $\sigma_1(\omega)$ $T=90$~K, 50~K, 35~K, and 20~K of
for $x = 0$ (a) and $x =0.4$ (b).
(c) and (d): dependence of $\sigma_1(\omega)$ on Br concentration for lowest $T=20$~K ($x$ = 0 and 0.4) and $T=5$~K ($x$ = 0.73, 0.85, and 0.9) (c) and $T=50$~K (d).}
\end{figure}

As demonstrated in Fig.~\ref{fig:gapa}(a), the pristine \etcl\
crystal shows typical signatures of a Mott insulator with an energy
gap gradually growing as the temperature drops below 50~K. There is
no clear-cut definition of the gap value: the extrapolation to zero
conductivity seems to be arbitrary. The fit of the conductivity
spectrum by a Lorentz oscillator leads to frequencies which are much
too high. An alternative way is to consider the energy range where
the spectral weight is low. If we choose the frequency $\Delta$ for
which the spectral weight reaches
$\int_0^{\Delta}\sigma_1(\omega){\rm d}\omega =
50000~\Omega^{-1}{\rm cm}^{-2}$, i.e.\ about 10\% of the total
spectral weight, we obtain a 50~K-gap value of $\Delta/2\pi c=
570$~\cm. As already pointed out by Kornelsen {\it et
al.},\cite{Kornelsen92} the gap grows almost linearly with
decreasing temperature (by about 10~\cm/K) indicating a second-order
transition. At 20~K, we obtain $\Delta/2\pi c= 900$~\cm and for
$T\rightarrow 0$ we extrapolate to about 1000~\cm; this is in good
agreement with the previous estimates. The result suggests that for
increasing temperature the carriers in \etcl\ are less localized.
The gap diminishes continuously with rising temperature as expected
for a mean-field transition, but does not disappear completely since
the material is a narrow-gap semiconductor at high temperatures.

For the sample with $x = 0.4$ (Fig.~\ref{fig:gapa}(b)) the
conductivity in the low-frequency region is still suppressed below
50~K, however, the gap value is significantly smaller than in the
pure Cl compound. Using the same method as above, we obtain
$\Delta/(2\pi c)= 400$~\cm\ at 50 K and $\Delta/(2\pi c)= 580$~\cm\
at 20 K, or for $T\rightarrow 0$ we extrapolate to about 650~\cm. In
agreement with the present observation, a considerable reduction of
the energy gap was recently predicted by cluster DMFT when $U/t$ is
reduced.\cite{Kyung06,Park08,Ohashi08} Most of the reduction is
attributed to the presence of two strong peaks in the spectral
functions at the gap edge induced by short-range antiferromagnetic
correlations in addition to the Hubbard bands. However, the broad
region of enhancement in the optical conductivity of
$\kappa$-(BE\-DT\--TTF)$_2$\-Cu\-[N\-(CN)$_{2}$]\-Br$_{0.4}$Cl$_{0.6}$
compared to the pristine Cl compound below 1250~\cm
(Fig.~\ref{fig:condsubtracted}(a)) indicates that these bands are
not as sharp as anticipated by the theoretical calculations.

The complete picture at the lowest temperature is demonstrated in
Fig.~\ref{fig:gapa}(c) where the conductivity of \etbrcl\ is plotted
for $x=0$, 0.4, 0.73, 0.85, and 0.9. When the Br concentration
increases above  $x = 0.4$, the optical gap is not substantially
reduced further but gradually filled in; a strong Drude contribution
develops. Indeed, with larger Br substitution, the relative Coulomb
repulsion $U/t$ is reduced. As it gets below a critical point
(between $x=0.4$ and 0.73), a  first-order transition occurs, where
the density of states at the Fermi level is supposed to rise
abruptly to a finite value.\cite{Bulla01} The coherent
quasiparticles form a Drude-like zero-frequency peak; this
contribution will be further analyzed in the following
Sec.~\ref{sec:Drude}.

The panel (d) of Fig.~\ref{fig:gapa} focusses on the optical
conductivity of \etbrcl\ at $T=50$~K, which is slightly above the
Mott-insulating or coherence temperature, respectively.
Accordingly, both the Mott gap in the low Br-doped compounds and
the Drude contribution in the salts with high Br content, are
suppressed. However, the conductivity below 500~\cm\ is still much
higher for the metallic compounds with $x=0.85$ and 0.9, showing a
wide peak at about 300~\cm\ as discussed in the previous section.

\subsubsection{Coherent quasiparticle response}
\label{sec:Drude} According to the Drude model, the optical response
of the free carriers in a conventional metal is restricted to a
zero-frequency mode. Its spectral weight $\int\sigma_1(\omega)\,{\rm
d}\omega$ is temperature independent. As the temperature is lowered
and phonon scattering freezes out, the Drude peak becomes narrower
and $\sigma(T)$ increases accordingly. Contrary, in heavy fermions
which are benchmark systems for the physics of strongly-correlated
electrons the spectral weight of the zero-frequency contribution
typically condenses when $T$ drops below some coherence temperature,
because the effective mass $m^*(T)$ increases due to
electron-electron interactions.\cite{Degiorgi99} At a
metal-insulator transition (caused by electronic correlations or
other reasons) the Drude spectral weight abruptly vanishes
$D(T)\rightarrow 0$.
\begin{figure}
\includegraphics[width=75mm]{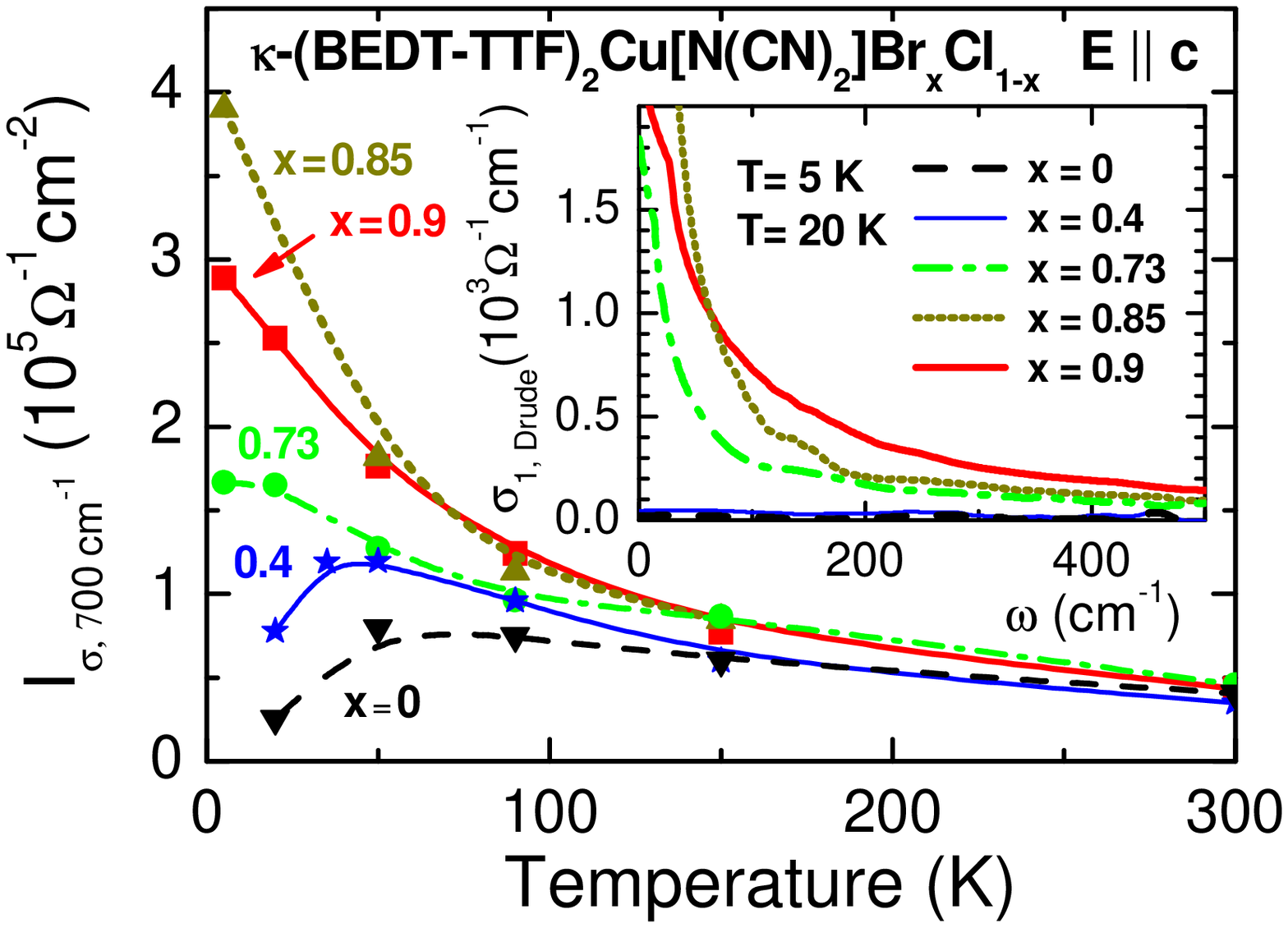}
\caption{\label{fig:Drudeweight}
(Color online) Temperature dependence of the spectral weight $I_{\sigma}$ of the conduction
electrons in  \etbrcl\ with $x=0$, 0.4, 0.73, 0.85, and
0.9 calculated via integration of $\sigma_1$ plotted in Fig.~\ref{fig:condsubtracted} up to $\omega_c/(2\pi c)=700$~\cm\ for $E\parallel c$. The lines are guides to the eye.
The inset shows the Drude conductivity $\sigma_{1,{\rm Drude}}$ obtained after subtraction of all finite-frequency modes at lowest temperatures ($T=20$~K for $x$ = 0 and 0.4 and $T=5$~K for $x$ = 0.73, 0.85, and 0.9).
}
\end{figure}

In order to get information on the correlation and temperature
evolution of the coherent quasiparticle response in \etbrcl\ we
calculate the spectral weight of the low-frequency part of the
spectra shown in Fig.~\ref{fig:condsubtracted}. Since the spectra
exhibit a minimum in the range between 500 and 800~\cm\ for any $x
\ge 0.73$ at low temperatures [Fig.~\ref{fig:condsubtracted}(a)], we
have picked $\omega_c/(2\pi c)= 700$~\cm\ as a possible cut-off
frequency. The temperature dependence of $I_{\sigma,
\omega_c=700~{\rm cm}^{-1}}$ is plotted in
Fig.~\ref{fig:Drudeweight}. For all values of $x$, the intensity of
the low-frequency spectral weight increases as the temperature is
lowered from room temperature down to 50~K. Reducing $T$ even
further, $I_{\sigma, \omega_c=700~{\rm cm}^{-1}}$ drops
significantly for the $x=0$ and 0.4 samples because the Mott-Hubbard
gap opens as the insulating state is entered. As the Drude peak
develops at $T\le50$~K, the metallic samples ($x \ge$ 0.73) exhibit
a steady enhancement $I_{\sigma, \omega_c=700~{\rm cm}^{-1}}$, the
effect is strongest in the samples containing the highest amount of
Br.

In an alternative approach, the zero-frequency mode can be
disentangled from the transitions between the Hubbard bands by
subtracting the latter. It is obvious from
Fig.~\ref{fig:condsubtracted} that this is only possible at lowest
temperatures where both contributions are well separated. The
far-infrared spectral range of the resulting Drude conductivity
$\sigma_{1,{\rm Drude}}$ is shown in the inset of
Fig.~\ref{fig:Drudeweight}. Interestingly, the width of the Drude
peak is much smaller for $x = 0.85\%$ compared to $x = 0.9\%$ while
the amplitude, i.e., the dc conductivity, is much higher in the
former.\cite{remark5}

\begin{figure}
\includegraphics[width=55mm]{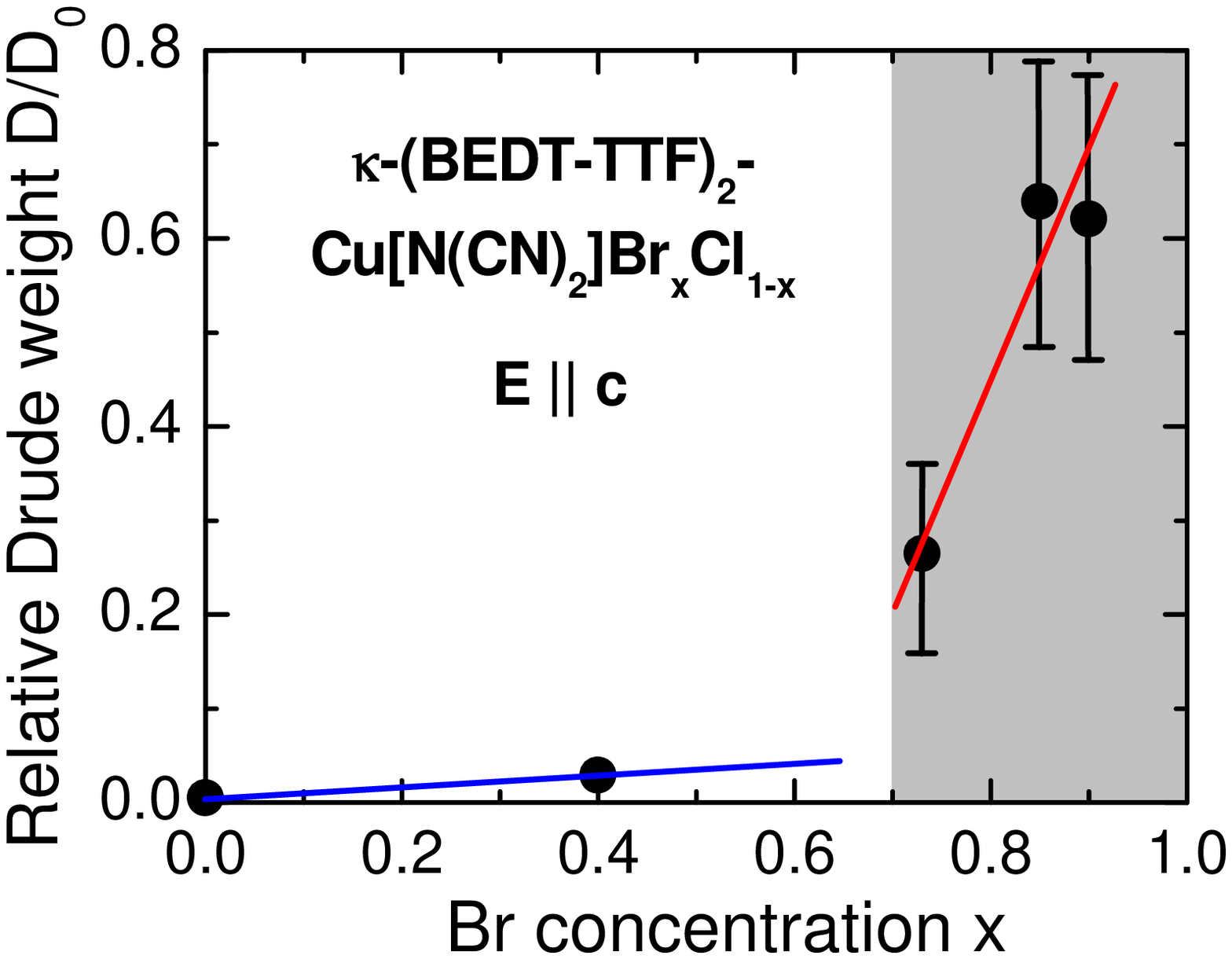}
 \caption{\label{fig:Drude}(Color online)
Ratio of the low-temperature Drude spectral weight $D(x)$ to the total spectral weight of the correlated charge carriers $D_0(x)$
at 20 K ($x=0$ and 0.4) and $T =5$~K ($x=0.73$, 0.85, 0.9).  The solid lines are guides to the eyes and
help to identify the Mott-insulating and metallic regimes (grey area) with a boundary around
$x_c=0.7$.}
\end{figure}

The strength of the coherent quasiparticle response can be estimated
from the relative Drude weight $D/D_0$ (Fig.~\ref{fig:Drude}) where
$D$ is the spectral weight $I_{\sigma}$ of the quasiparticle peak
obtained by integration of the data shown in the inset of
Fig.~\ref{fig:Drudeweight} and $D_0$ the total spectral weight of
the coherent carriers shown in Fig.~\ref{fig:Neff}. From the
simplest linear interpolation, two regimes can be identified at
lowest temperature as illustrated in Fig.~\ref{fig:Drude}. A Drude
response is found only for $x>0.7$, it rapidly grows in weight as
the Br content increases, i.e., the relative strength of the Coulomb
interaction $U/t$ decreases. The ratio $D/D_0$, should exhibit a
jump at $(U/t)_c$.\cite{Carpone01,Powell05} The limited number of Br
concentrations $x$ available for our study does not allow us to give
a definite number for $D/D_0$ at the critical point. Since the phase
transition at the critical $(U/t)_c$ is of first order, Bulla {\em
et al.}\cite{Bulla01} predicted that the abrupt change of the Drude
weight $D(x)$ even shows some hysteresis which might be checked in
the future by very precise far-infrared spectroscopy with continuous
variation of the external pressure. Actually, in transport
measurements as function of pressure a hysteresis between increasing
and decreasing pressure sweeps was reported.\cite{LIM03,KAG04}

\subsubsection{Suppression of the spectral weight across the transition}
The optical-spectral-weight redistribution can be analyzed more
precisely based on the restricted f-sum rule for the effective
one-band Hubbard model:\cite{Maldague77}
\begin{equation}
\int_0^\infty \sigma_1(\omega)\, {\rm d}\omega= {\pi N \over 2 D}
{e^2 d^2 \over \hbar^2 M \Omega} \langle -E_{\rm kin} \rangle
\quad , \label{res_fsum}
\end{equation}
where $D$ is the dimension and $N$ is the total number of carriers
per dimer. $M$ is the number of dimers in the crystal, $\Omega$ the
volume per (BEDT-TTF)$_2$ dimer and $d$ the lattice parameter.
Eq.~(\ref{res_fsum}) brings a dynamical quantity, which probes the
optical transitions in the system, together with a ground-state
quantity, the total kinetic energy of the many-body system (the
Hubbard model) and depends on temperature $T$ and Coulomb repulsion
$U$ in contrast to the full sum rule given in Eq.~(\ref{fsum}).

As we already have shown in Ref.~\onlinecite{Merino08}, the spectral
weight of the system is indeed suppressed: while for a
non-interacting system the DMFT calculation gives a result for the
spectral weight of ${ N \over 4 D} {e^2 d^2 \over \hbar^2 c_0 M
\Omega} \langle -E_{\rm kin} \rangle_0= 9.16\cdot10^5~\Omega^{-1}$
cm$^{-2}$, the calculated spectral weight of a system with
$U=0.3$~eV  is $5.8 \cdot10^5~\Omega^{-1}{\rm cm}^{-2}$  in good
agreement with the experimental result $6.4
\cdot10^5~\Omega^{-1}{\rm cm}^{-2}$ obtained for
$\kappa$-(BE\-DT\--TTF)$_2$\-Cu\-[N\-(CN)$_{2}$]\-Br$_{0.73}$Cl$_{0.27}$
with a cut-off frequency in the order of $U$. A similar effect of
suppressed $E_{\rm kin}$ is observed in optical experiments on
cuprate superconductors \cite{Uchida91,Padilla05} which are
considered to be doped Mott insulators.

\begin{figure}
\includegraphics[width=50mm]{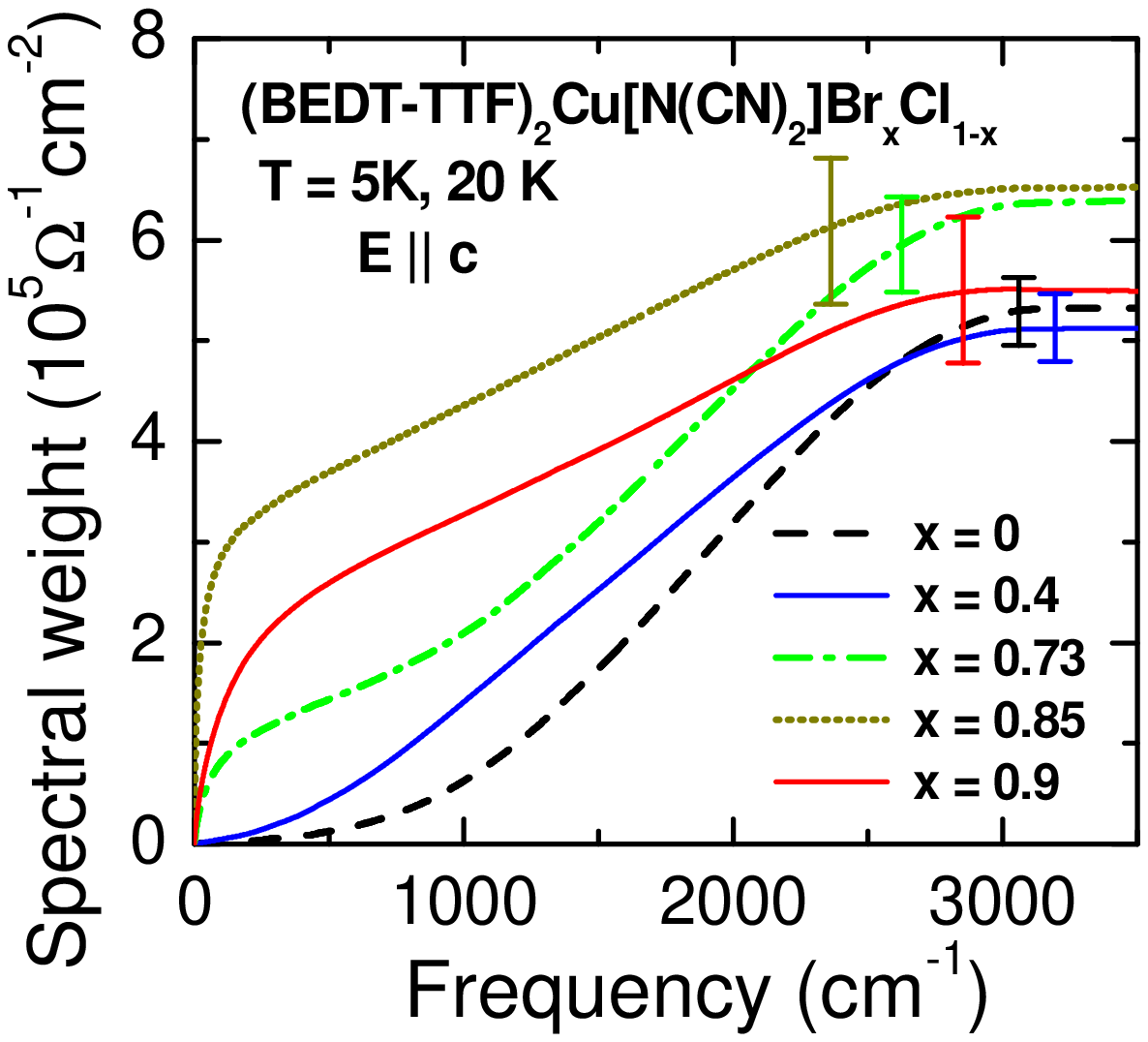}
\caption{(Color online)
Spectral-weight distribution
for $\kappa$-(BEDT-TTF)$_2$Cu[N(CN)$_2$]Br$_x$Cl$_{1-x}$ at low temperature. The spectral weight is redistributed
as one approaches the transition to the Mott-insulating phase.
Typical error bars for the experimental data are shown.\cite{remark5} \label{fig:Neff}}
\end{figure}
In Fig.~\ref{fig:Neff} we present for the first time the development
of the spectral weight of the correlated charge carriers  across the
metal-insulator transition at low temperatures. For all compounds,
at frequencies of the order of $U$ [i.e., $\omega/(2\pi c) \gtrsim
3000$ cm$^{-1}$] the spectral weight saturates. Due to the strong
electronic correlations it is considerably reduced;\cite{Merino08}
this effect gets stronger with increasing correlations and for the
insulating compounds $x<0.73$ a further suppression is observed.
Here, the contribution of the zero-frequency mode to the spectral
weight is missing; parts of the spectral weight are transferred to
higher frequencies - to the Hubbard band which has more than twice
the amplitude for $x=0$ compared to $x =$ 0.85 and 0.9 (see
Fig.~\ref{fig:condsubtracted}). However, not all the spectral weight
is recovered in the Hubbard bands, but is reduced in the
high-$\omega$ limit compared to the metallic compounds.

\subsection{Electronic correlations in the vicinity of the Mott transition}
\label{sec:mass}
 In the analysis of the electrodynamics of the
correlated charge carriers in \etbrcl\ given above we used a
multicomponent model where we separated a free-carriers
zero-frequency mode and a finite-frequency mode due to transitions
between Hubbard bands. However, for the low-temperature state of the
metallic compounds with $x = 0.73$, 0.85, and 0.9, where the
zero-frequency peak is strong and overlaps with the finite-frequency
contributions, we may use an alternative description by a
one-component model. Here, only itinerant charge carriers account
for the frequency dependence of the optical conductivity shown in
Fig.~\ref{fig:condsubtracted}(a). The same dual approach was
utilized to describe the in-plane electromagnetic response of the
high-$T_c$ cuprates.\cite{Basov05} The multicomponent model gives a
reasonable description for the strongly-underdoped cuprates, while
for optimally and overdoped cuprates the one-component model seems
to be more appropriate. \cite{Basov05, Lee05} From the conductivity
spectra we can extract the frequency dependence of scattering rate
$\Gamma_1(T,\omega)$ and renormalized mass of the charge carriers
which is, as well as the temperature dependence of both, governed by
many-body effects.

A fingerprint of strong electron-electron interactions is the $T^2$
dependence of the scattering rate and consequentially of the dc
resistivity.  For example, in the resistivity of heavy fermions a
$T^2$ behavior was observed, however, there the experimental
findings were limited to very low temperatures and rather small
temperature intervals.\cite{Stewart84,Degiorgi99} An example of a
two-dimensional Fermi liquid where the in-plane and out-of-plane
resistivity follows a $T^2$ law is Sr$_2$RuO$_4$. \cite{Maeno97}
Transport measurements of the in-plane or out-of-plane resistivity
of metallic \etbr\ or \etcl\ with applied hydrostatic pressure also
indicated $\rho \propto T^2$ in the temperature region between the
superconducting transition and approximately 40
K.\cite{Dressel93,Dressel96,LIM03,Strack05,Faltermeier07} that was
interpreted in terms of the Fermi-liquid model.\cite{Merino00} But
this conclusion was always under debate since the temperature region
is very limited and other models also propose a quadratic
temperature dependence of the resistivity.
\cite{Stewart84,Weger95,Strack05}

Optical spectroscopy offers an alternative way to probe
the electron-electron interactions. According to Fermi-liquid theory
\begin{equation}
\Gamma_1(T,\omega)= A\left[ (2\pi k_B T)^2 +
(\hbar\omega)^2\right] \quad ;\label{eq:Fermi}
\end{equation}
the $T^2$ dependence of the scattering rate $\Gamma_1=1/\tau$ should
be accompanied by a similar parabolic frequency dependence. To
extract information about the frequency-dependent scattering rate
$\Gamma_1(T,\omega)$ and effective mass, we analyze the
low-temperature spectra using a generalized Drude
model;\cite{Allen77,DresselGruner02} this approach is commonly
applied to correlated-electron systems like heavy fermions and
high-temperature
superconductors:\cite{Degiorgi99,Dressel02,Dordevic06,Basov05}
\begin{equation}
\hat{\sigma}(\omega)=\frac{\omega_p^2}{4\pi}
\frac{1}{\Gamma_1(\omega)-i\omega(m^*(\omega)/m_{\rm b,opt})} \quad
. \label{eq:gen-drude2}
\end{equation}
Here a $\Gamma_1(\omega)$ is the real part of the complex scattering
rate $\hat{\Gamma}(\omega)=\Gamma_1(\omega)+i\Gamma_2(\omega)$, with
the imaginary part related to the enhanced mass (renormalized due to
electron-electron interactions) $m^*/m_{\rm
b,opt}=1-\Gamma_2(\omega)/\omega$. From the complex conductivity we
obtain expressions for $\Gamma_1(\omega)$ and $m^*(\omega)$ in terms
of $\sigma_1(\omega)$ and $\sigma_2(\omega)$ as follows:
\begin{equation}
\Gamma_1(\omega)=\frac{\omega_p^2}{4\pi}
\frac{\sigma_1(\omega)}{|\hat{\sigma}(\omega)|^2}\quad ;\quad
\frac{m^*(\omega)}{m_{\rm b,opt}}=\frac{\omega_p^2}{4\pi}
\frac{\sigma_2(\omega)/\omega}{|\hat{\sigma}(\omega)|^2}\quad.\label{eq:gam-w}
\end{equation}

\begin{figure}
\scalebox{0.45}{\includegraphics*{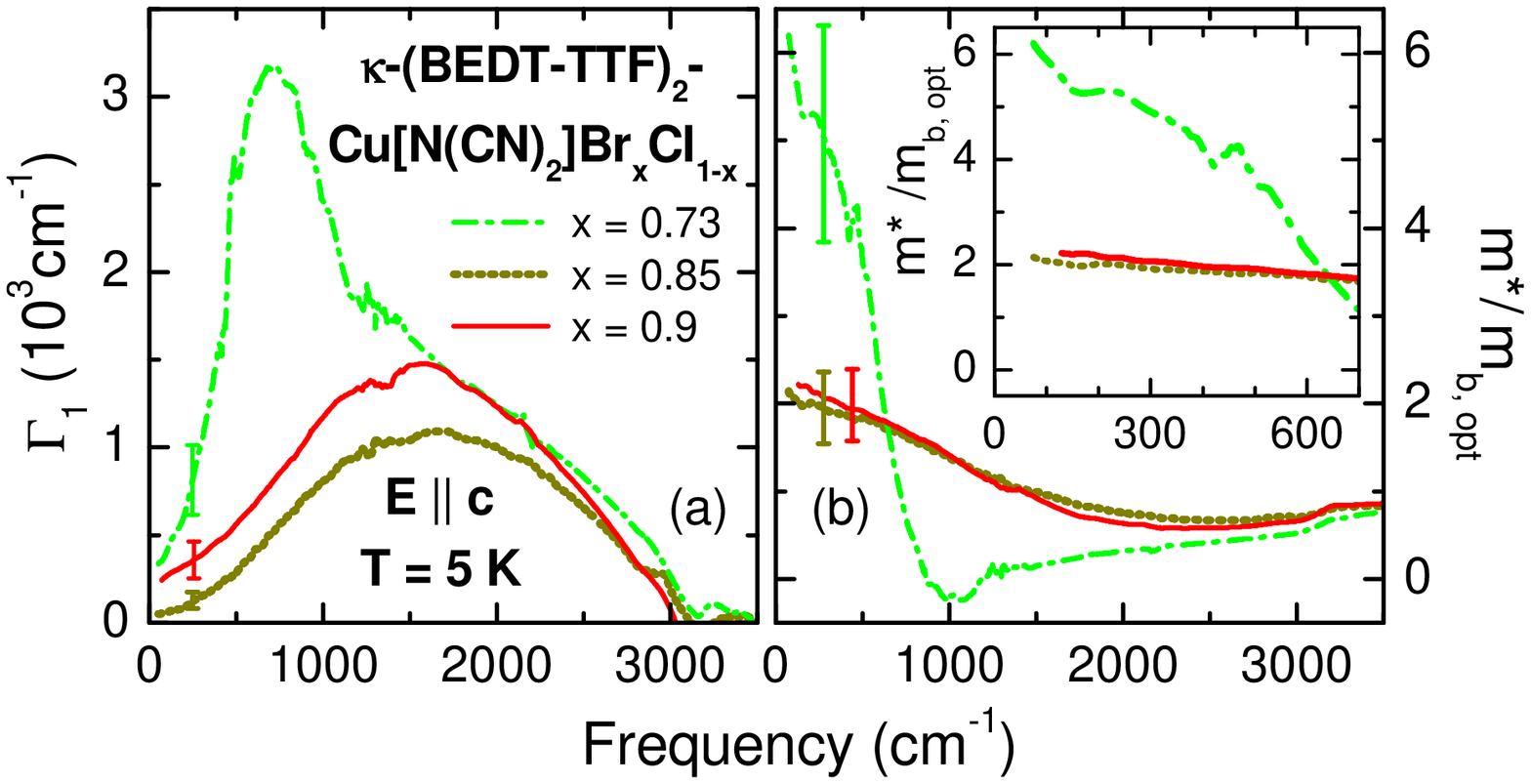}}
\caption{\label{fig:mass}(Color online) Frequency dependence of the
low-temperature ($T=5$~K) scattering rate $\Gamma_1(\omega)$
(left panel) and effective mass $m^*/m_{b,\rm opt}$ (right panel and inset) of
\etbrcl\ for different $x$ as indicated. Typical error bars for the experimental data are shown.}
\end{figure}
Fig.~\ref{fig:mass} shows $\Gamma_1(\omega)$ and
$m^*(\omega)/m_{b,\rm opt}$ for \etbrcl\ samples with Br
concentrations $x \ge 0.73$. The scattering rate $\Gamma_1$ consists
out of a frequency-independent part $\Gamma_1 (0)$, which is derived
from the limit $\omega \rightarrow 0$, and a frequency-dependent
part. We obtained $\Gamma_1(0) = 315$ \cm\ ($x=0.73$), 48 \cm\
($x=0.85$), and 280 \cm\ ($x=0.9$) at $T = 5$~K.\cite{remark9} The
frequency dependencies of the scattering rate and effective mass in
the less correlated compounds with $x = 0.85$ and 0.9 have similar
smooth characteristics as expected for a small variation in doping;
$\Gamma_1(\omega)$ increases up to the maximum at a frequency
$\omega_m$ and then decreases towards higher frequencies.
$\kappa$-(BEDT-TTF)$_2$\-Cu\-[N\-(CN)$_{2}$]\-Br$_{0.73}$Cl$_{0.27}$,
located very close to the metal-insulator transition, has a
substantially enhanced scattering rate and effective mass with very
strong frequency dependencies.  We find the maximum in
$\Gamma_1(\omega)$ that corresponds to a drop in
$m^*(\omega)/m_{b,\rm opt}(\omega)$ at $\omega_m/(2\pi c) \approx
700$ \cm\ for $x = 0.73$ and $\omega_m/(2\pi c) \approx 1600$ \cm\
for $x = 0.85$ and 0.9, respectively.  A presence of a peak in
$1/\tau(\omega)$ and a sharp drop in $m^*(\omega)$ at a certain
frequency $\omega_m$ is in excellent agreement with DMFT
calculations.\cite{Merino08}

\begin{figure}[t]
\scalebox{0.425}{\includegraphics*{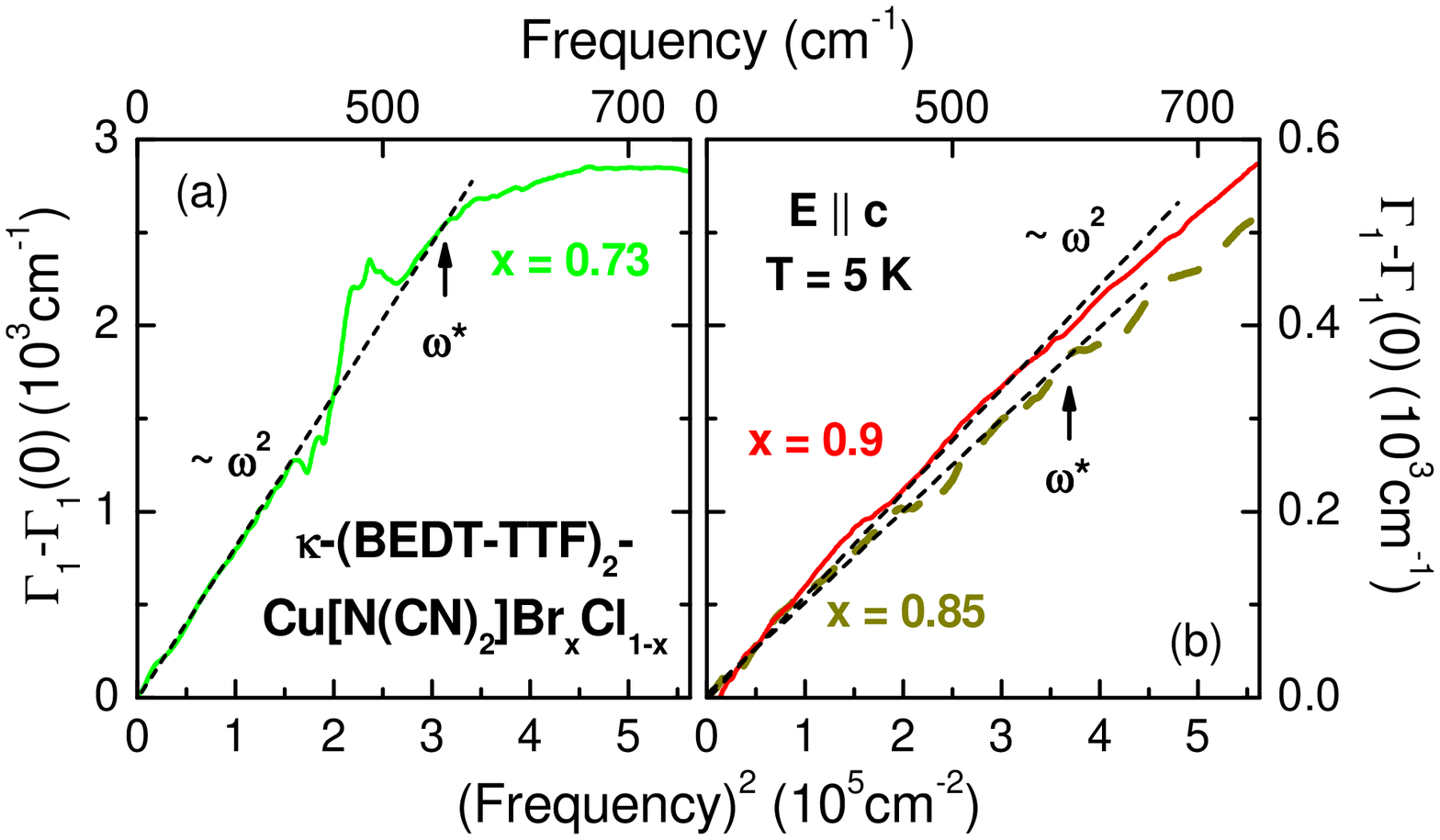}}
\caption{\label{fig:nusquare}(Color online) Frequency-dependent part of the scattering rate $\Gamma_1(\omega) - \Gamma_1(0)$ at $T=5$~K
of \etbrcl\ as function of the squared frequency. We determined for the frequency-independent dc limit of the scattering rate
$\Gamma_1(0)$ the following values : 315 \cm ($x=0.73$), 48 \cm ($x=0.85$), and 280 \cm ($x=0.9$). Note the different vertical scales for both frames.}
\end{figure}

The frequency-dependent part of $\Gamma_1(\omega)$ well below
$\omega_m$ is analyzed in Fig.~\ref{fig:nusquare}. All investigated
metallic compositions are linear in the
$\left[\Gamma_1-\Gamma_1(0)\right]$ vs. $\omega^2$ representation
below a frequency scale $\omega^{*}$, where $\omega^{*}/(2\pi c)
\approx 500$~\cm\ for $x=0.73$ and  $\omega^{*}/(2\pi c) \approx
600$~\cm\ for $x=0.85$ and 0.9. Deviations from the $\omega^2$
behavior of the scattering rate expected in Fermi liquids occur for
$\omega > \omega^*$, with $\hbar\omega^* \ll \epsilon_F$, the Fermi
energy. Thus, $\omega^*$ is a low energy scale which separates
conventional metallic behavior from unconventional non-Fermi-liquid
behavior at large frequencies. In agreement with our experimental
values, DMFT calculations predict  a red shift of $\omega^*$ and
$\omega_m$ with increasing correlations $U/t$; the parameters
mentioned in Sec.~\ref{sec:MottHubbardsystem} give a good
description for $x = 0.73$ and yield $\omega^*/(2\pi c) \approx
400$~\cm.\cite{Merino08} It should be mentioned that cluster DMFT
\cite{Park08} suggests when the phase border is approached from the
metallic side an even further reduced coherence scale and a
coexistence region of the metallic and insulating phases due to
short-range correlations in the anomalous metallic state with a
non-quadratic scattering rate. Since there are no indications of
such a behavior in the data presented in this work, we conclude that
the $U/t$ and $T/t$ range of our experiments does not cover the
coexistence region.

Generally, it is predicted by Brinkman-Rice theory
\cite{Brinkman70,Imada98} and DMFT calculations \cite{Georges96}
that the electronic correlations become stronger on approaching the
Mott transition from the metallic side. Accordingly, the effective
mass gets enhanced. Resonating-valence-bond theory of the
Hub\-bard-Heisenberg model predicts a gradual increase of $m^*$ for
values of effective repulsion $U/t$ not too close to the first-order
Mott transition and a strong increase very close to the transition.
\cite{Powell05}  We find a mass enhancement (inset of
Fig.~\ref{fig:mass}, $\omega^*/(2\pi c)< 300$~\cm) of about a factor
2 in the less correlated compounds and of about 5-6 in the material
located very close to the Mott transition, which is in accord with
the theoretical prediction.

The slope $a$ in the linear regions of $[\Gamma_1-\Gamma_1(0)]$
plotted vs. $\omega^2$ in Fig.~\ref{fig:nusquare} is proportional to
the prefactor $A$ in Eq.~(\ref{eq:Fermi}). In Fermi liquids, $A$ is
related to the mass enhancement by $A\propto (m^{*})^2$, which leads
to the well-known Kadowaki-Woods ratio $A/\gamma^2$ = const.\ via
$\gamma \propto m^{*}$, where $\gamma$ is the Sommerfeld
coefficient.\cite{Kadowaki86} The comparative mass enhancement in
the low-frequency limit $m^{*}(x=0.73)/m^{*}(x=0.85)\approx 2.75$
and the enhancement of the prefactor obtained from the scattering
rate $A(x=0.73)/A(x=0.85)\approx 8$ evidences the excellent
fulfilment of this relation. It should be noted, however, that
pressure-dependent dc measurements of the in-plane resistivity in
\etcl\ yield a prefactor of $T^2$ at low temperatures that changes
by a factor of three when pressure is increased from 300 bar to 1
kbar.\cite{LIM03} If we compare the data of the samples with
$x=0.85$ and $x=0.9$ in Figs.~\ref{fig:mass} and \ref{fig:nusquare},
it is obvious that $A$ and $m^{*}$ are about the same in both as
expected for a small variation in doping.\cite{remark9} According to
the relations $A\propto (m^{*})^2 \propto (T_F^{*})^{-2}$, the
prefactor $A$ and $m^*$ are also expected to scale with the
effective Fermi temperature $T_F^{*}$. \cite{Strack05} If we
associate $T_F^{*}$ with the energy scale below which we observe the
$\omega^2$ dependence, i.e.\ $\omega^*$, we cannot identify such a
scaling. Interestingly, the energy scale $\omega_m$, which marks the
maximum in $\Gamma_1(\omega)$, seems to fulfil this relation much
better.

The strong mass renormalization which we note in the
bandwidth-controlled Mott transition in \etbrcl\ is in contrast to
observations in two-dimensional doped Mott insulators like the
cuprates where the Mott transition is obtained by band-filling
control.\cite{Padilla05} There, the transition into the
Mott-insulating state is of the vanishing-carriers type and the
effective mass stays constant. An investigation of the evolution of
the effective mass close to the Mott transition in V$_{2-y}$O$_3$
confirmed that the effective mass diverged as function of
hydrostatic pressure while it slightly decreased as function of hole
doping $y$ if the metal-insulator transition was approached.
\cite{Carter93}

\section{Conclusion}
The polarized reflection spectra of \etbrcl\ have been
systematically investigated in a frequency range from the very
far-infrared to the near-infrared. This series of alloys is a
benchmark system for the bandwidth-controlled Mott transition, with
$U/t$ decreasing as function of Br content $x$. In the present study
we follow the correlation and temperature dependence of the
correlated-carriers dynamical response on both sides of the Mott
transition. We compare the findings with predictions of the
single-site and cluster DMFT calculations for a half-filled Hubbard
model.

At elevated temperatures $T \gg 50$~K, all compounds are narrow-gap
semiconductors and show no substantial differences in the optical
spectra. The Mott-Hubbard physics is confined to low temperatures $T
< 50$~K: in the Br-rich samples, the zero-frequency quasiparticle
peak accounts for the metallic conductivity. With increasing $U/t$,
the Drude spectral weight is suppressed and totally vanishes beyond
the first-order phase transition at $x_c\approx 0.7$ in accord with
theoretical predictions. In the Cl-rich Mott-insulating samples, an
energy gap opens and increases as the electronic correlations get
stronger and reaches almost 1000~\cm\ for $T\rightarrow 0$ in the
pristine Cl compound. The enhanced $U/t$ thus results in a
redistribution and suppression of the spectral weight of the
correlated charge carriers, including the transitions between the
Hubbard bands and, if present, the Drude weight. Below a
characteristic frequency $\omega^*/(2\pi c)\approx 500-600$~\cm, the
quasiparticles in the metallic phase show typical signatures of a
Fermi liquid: a $A\omega^2$ dependence of the scattering rate
$\Gamma_1(\omega)$ and a substantial enhancement of the effective
mass $m^{*}(\omega)$, where the prefactor $A$ and
$m^{*}(\omega\rightarrow 0)$ follow the $A\propto (m^{*})^2$
scaling.  Both parameters diverge as $x_c$ is approached. Above
$\omega^*$, a transition to an unconventional non-Fermi liquid
regime is observed.

\acknowledgements We thank J. Merino and R. McKenzie for helpful discussions of theoretical aspects.
 The project was supported by the
Deutsche Forschungsgemeinschaft. ND is grateful for the support by
the Alexander von Humboldt-Foundation and by the
Magarete-von-Wrangell-Programm of Baden-W\"urttemberg.


\begin{thebibliography}{99}
\bibitem{Mott74}N.F. Mott, {\em Metal-Insulator Transitions} (Taylor \&\ Francis, London, 1974); F. Gebhard, {\em The Mott Metal-Insulator Transition} (Springer, Berlin, 1997).
\bibitem{Fulde93}P. Fulde, {\em Electron Correlations in Molecules and
Solids}, 3rd edition (Springer-Verlag, Berlin, 2002).
\bibitem{Grewe91}N. Grewe and F. Steglich, in: Handbook on the Physics and Chemistry
 of Rare Earths, V. {\bf14}, ed.: K.A. Gscheidner Jr. and L. Eyring
 (Elsevier, Amsterdam - New York, 1991), p. 343.
\bibitem{Imada98}
 M. Imada, A. Fujimori, and Y. Takura, Rev. Mod. Phys. {\bf 63},  1 (1998).
\bibitem{Jerome94}D. J\'erome,  in {\em Organic Conductors}, edited by J.-P. Farges (Marcel Dekker, New York, 1994), p.\ 405.
\bibitem{Ishiguro98}T. Ishiguro, K. Yamaji, and G. Saito, {\em Organic
Superconductors}, 2nd edition (Springer-Verlag, Berlin, 1998).
\bibitem{McKenzie98} R.H. McKenzie, Comments Cond.\ Mat.\ {\bf 18}, 309 (1998).
\bibitem{Seo04}H. Seo, C. Hotta, and H. Fukuyama, Chem. Rev. {\bf 104},
5005 (2004).
\bibitem{Dressel04}
M. Dressel and N. Drichko, Chem. Rev. {\bf 104}, 5689 (2004).
\bibitem{Fukuyama06}
H. Fukuyama, J. Phys. Soc. Jpn. {\bf 75}, 051001 (2006).
\bibitem{remark0}BEDT-TTF stands for
bis-(ethyl\-ene\-di\-thio)\-te\-tra\-thia\-ful\-va\-lene.
\bibitem{Lefebvre00}S. Lefebvre, P. Wzietek, S. Brown, C. Bourbonnais,
D. J\'erome, C. M\'ezi\`ere, M. Fourmigu\'e, and P. Batail, Phys. Rev. Lett. {\bf 85}, 5420 (2000).
\bibitem{LIM03}P. Limelette, P. Wzietek, S. Florens, A. Georges, T.A. Costi,
C. Pasquier, D. Jerome, C. M\'eziere, and P. Batail, Phys. Rev. Lett.
{\bf 91}, 016401 (2003).
\bibitem{KAG04}F. Kagawa, T. Itou, K. Miyagawa, and K. Kanoda, Phys. Rev. B {\bf 69}, 064511 (2004); F. Kagawa, T. Itou, K. Miyagawa, and K. Kanoda, Phys. Rev. Lett. {\bf 93}, 127001 (2004); F. Kagawa, K. Miyagawa, and K. Kanoda, Nature {\bf 436}, 534 (2005).
\bibitem{Georges96} A. Georges, G. Kotliar, W. Krauth,  and
M.J. Rozenberg, Rev. Mod. Phys. {\bf  68}, 13 (1996).
\bibitem{Kotliar04}G. Kotliar and D. Vollhardt, Physics Today, March  2004, p.~53.
\bibitem{Kino96}H. Kino and H. Fukuyama, J. Phys. Soc. Jpn. {\bf 64}, 2726 (1995); {\it ibid.} {\bf 64}, 4523 (1995); {\it ibid.} {\bf 65}, 2158 (1996).
\bibitem{Merino00}J. Merino and R. H. McKenzie, Phys. Rev. B {\bf 61}, 7996 (2000); {\it ibid.} {\bf 62}, 16442 (2000).
\bibitem{Merino08}
J. Merino, M. Dumm, N. Drichko, M. Dressel, and R.H. McKenzie,
Phys. Rev. Lett. {\bf 100}, 086404 (2008).
\bibitem{Eldridge91b}
J.~E. Eldridge, K. Kornelsen, H.~H. Wang, J.~M. Williams, A.~V.~D.
Crouch, and D.~M. Watkins, Solid State Commun. {\bf 79},  583
(1991).
\bibitem{Kornelsen92}
K. Kornelsen, J.~E. Eldridge, H.~H. Wang, H.~A. Charlier, and J.~M.
Williams,  Solid State Commun. {\bf 81},  343  (1992).
\bibitem{Haas00}P. Haas, E. Griesshaber, B. Gorshunov,  D. Schweitzer, M. Dressel, T. Klausa,  W. Strunz, and F. F. Assaad,
Phys. Rev. B {\bf 62}, R14673 (2000).
\bibitem{Dressel03}M. Dressel, N. Drichko, J. Schlueter, and J. Merino, Phys. Rev. Lett. {\bf 90}, 167002 (2003).
\bibitem{Basov05} D.N. Basov and T. Timusk, Rev. Mod. Phys.
{\bf 77}, 721 (2005).
\bibitem{Drichko06}N. Drichko, M. Dressel, C. A. Kuntscher, A. Pashkin, A. Greco, J. Merino, and J. Schlueter
Phys. Rev. B {\bf 74}, 235121 (2006).
\bibitem{Sasaki08}T. Sasaki, N. Yoneyama, Y. Nakamura, N. Kobayashi, Y. Ikemoto, T. Moriwaki, and H.
Kimura, Phys. Rev. Lett. {\bf 101}, 206403 (2008).
\bibitem{Faltermeier07}
D. Faltermeier, J. Barz, M. Dumm, M. Dressel, N. Drichko, B. Petrov,
V. Semkin, R. Vlasova, C. M\'eziere, and P. Batail, Phys. Rev. B
{\bf 76}, 165113 (2007).
\bibitem{Rice76}M.J. Rice, Phys. Rev. Lett. {\bf 37}, 36 (1976).
\bibitem{Yartsev93}Delhaes P., Yartsev V.M., Advances in Spectroscopy,  {\bf 22}, R.J.H. Clark, R.E. Hester Eds. {\it John Wiley and Sons,
1993}, p.~199; V. M. Yartsev, O. Fichet, J.-P. Borgion and P.
Delhaes. J. Phys. II France {\bf 3}, 647 (1993)
\bibitem{YAR96}V.M Yartsev , O.O. Drozdova, V.N. Semkin and R.M. Vlasova,  J. Phys. I (France) {\bf 6}, 1673 (1996);
V.M. Yartsev, in: {\em Materials and Measurements in Molecular
Electronics}, ed. by K. Kajimura and S. Kanoda (Springer-Verlag,
Berlin 1996), p. 189; V. M. Yartsev and A. Graja, Int. Journ. of
Mod. Phys. B {\bf 12}, 1643 (1998).
\bibitem{Dressel09}M. Dressel D. Faltermeier, M. Dumm, N. Drichko,
B. Petrov, V. Semkin, R. Vlasova,  C. M\'eziere, and P. Batail,
Physica B {\bf 404}, 541 (2009).
\bibitem{remark1}Here we should note that
the substitution of Cl
by Br in the anion layers does not simply and homogeneouly reduce
the  lattice parameters. In the table below the unit cell parameters
and volume of \etcl\ and \etbr\ are determined at room temperature.
The overlap integrals $t$ are given at
$T=127$~K.\protect\cite{Mori99,Seo04}\newline
\begin{tabular}{lcc}
\hline\hline
         &  \multicolumn{2}{c}{$\kappa$-(BEDT-TTF)$_2$Cu[N(CN)$_2$]$X$}\\
     $X$ & Cl & Br \\
     \hline
$a$ (\AA) & 12.977 & 12.942 \\
$b$ (\AA) & 29.979 & 30.016 \\
$c$ (\AA) &  8.480 &  8.539 \\
$V$ (\AA$^3$) & 3299 & 3317 \\
$t_1$ (meV) &  26 &  25 \\
$t_2$ (meV) &  36 &  38 \\
\hline\hline
\end{tabular}
\newline
In fact while $a$ decreases when going from \etcl\ to \etbr, the $c$
axis increases, leading to stronger frustration for \etbr. This
effect might be of superior importance for formation of the coherent
particles.
\bibitem{DresselGruner02}M.~Dressel and G.~Gr\"{u}ner, {\it Electrodynamics of Solids}  (Cambridge University Press, Cambridge, 2002).
\bibitem{remark2}At $T = 300$~K, the spectral weight is not
completely recovered in most of the compositions. At least parts of
the missing weight can be attributed to the subtantial thermal
expansion of the crystals. When heated up from 20~K to 295~K, the
volume of \etbr\ increses by about $3\%$.\protect\cite{Watanabe91}
\bibitem{Rozenberg95} M.J. Rozenberg, G. Kotliar, H. Kajueter,
G.A. Thomas, D.H. Rapkine, J.M. Honig, and P. Metcalf, Phys. Rev.
Lett. {\bf 75}, 105 (1995).
\bibitem{Kyung06}B. Kyung, S.S. Kancharla, D. S\'en\'echal, A.-M. S. Tremblay, M. Civelli, and G. Kotliar, Phys. Rev. B {\bf 73}, 165114 (2006);
B. Kyung and A.-M. S. Tremblay, Phys. Rev. Lett. {\bf 97}, 046402
(2006).
\bibitem{Park08}H. Park, K. Haule, and G. Kotliar, Phys. Rev. Lett. {\bf 101},
186403 (2008).
\bibitem{Ohashi08}T. Ohashi, T. Momoi, H. Tsunetsugu, and N. Kawakami, Phys. Rev. Lett. {\bf 100}, 076402 (2008).
\bibitem{remark7}Park {\it et al.} call it a ``bad insulator''.\protect\cite{Park08}
\bibitem{Bulla01}R. Bulla, T.A. Costi, and D. Vollhardt, Phys. Rev. B {\bf 64}, 045103 (2001).
\bibitem{Degiorgi99}L. Degiorgi, Rev. Mod. Phys. {\bf 71}, 687
(1999).
\bibitem{remark5}In the analysis given in Section \ref{sec:mass} we will
attribute the discrepancy between the $x=0.85$ and 0.9 data to the
different frequency-independent parts of the scattering rate
$\Gamma_1(0)$ in both compounds (see also
Ref.~\onlinecite{remark9}).
\bibitem{Carpone01}M. Capone, L.
Capriotti, F. Becca, S. Caprara, Phys. Rev. B {\bf 63}, 085104
(2001).
\bibitem{Powell05}B.J. Powell and R.H.
McKenzie, Phys. Rev. Lett. {\bf 94},
047004 (2005).
\bibitem{Maldague77}
P.F. Maldague, Phys. Rev. B {\bf 16}, 2437 (1977).
\bibitem{Uchida91}
 S. Uchida, T. Ido, H. Takagi, T. Arima, Y. Tokura and S. Tajima, Phys. Rev. B {\bf 43}, 7942 (1991).
\bibitem{Padilla05}
 W.J. Padilla,  Y.S. Lee, M. Dumm, G. Blumberg, S. Ono, Kouji Segawa, Seiki Komiya, Yoichi Ando, D.N. Basov, Phys. Rev. B {\bf 72}, 060511(R) (2005).
\bibitem{Lee05}Y.S. Lee {\it et al.}, Phys. Rev. B {\bf 72}, 054529 (2005).
\bibitem{Stewart84}G.R. Stewart, Rev. Mod. Phys. {\bf 56}, 755 (1984).
\bibitem{Maeno97}Y. Maeno {\it et al.}, J. Phys. Soc. Jpn. {\bf 66}, 1405 (1997).
\bibitem{Dressel93}M. Dressel, S. Bruder, G. Gr\"{u}ner, K.D. Carlson, H.H. Wang, and J.M.
Williams, Phys. Rev. B {\bf 48}, 9906 (1993); M. Dressel, O. Klein,
G. Gr\"{u}ner, K.D. Carlson, H.H. Wang, and J.M. Williams, Phys. Rev. B
{\bf 50}, 13603  (1994).
\bibitem{Dressel96}M. Dressel and G. Gr\"uner, Mol. Cryst. Liq.
Cryst. {\bf 248}, 107 (1996).
\bibitem{Strack05}Ch. Strack, C. Akinci, V. Paschenko, B. Wolf, E. Uhrig, W. Assmus, M. Lang, J. Schreuer, L. Wiehl, J.A. Schlueter, J. Wosnitza, D. Schweitzer,
J. M\"uller, and J. Wykhoff, Phys. Rev. B {\bf 72}, 054511 (2005).
\bibitem{Weger95}M. Weger and D. Schweitzer, Synth. Met.  {\bf 70}, 889 (1995); J. Hagel, J. Wosnitza, C. Pfleiderer, J. A. Schlueter,
J. Mohtasham, and G. L. Gard, Phys. Rev. B {\bf 68}, 104504 (2003).
\bibitem{Allen77}J.W. Allen and J.C. Mikkelsen, Phys. Rev. B {\bf 15}, 2952 (1977).
\bibitem{Dressel02}M. Dressel, N.V. Kasper, K. Petukhov, B. Gorshunov, G. Gr\"uner, M. Huth, and H. Adrian, Phys. Rev. Lett. {\bf 88}, 186404 (2002);
M. Dressel, N.V. Kasper, K. Petukhov, D.N. Peligrad, B. Gorshunov,
M. Jourdan, M. Huth, and H. Adrian, Phys. Rev. B {\bf 66}, 035110
(2002).
\bibitem{Dordevic06}S.V. Dordevic and D.N. Basov, Ann. Physik {\bf 15}, 545 (2006).
\bibitem{remark9} The enhancement of $\Gamma_1(0)$ by a factor of about
6 observed in the $x = 0.9$ sample (compared to $x=0.85$) might
originate from the presence of more crystal imperfections. Indeed,
the surface of the $x = 0.9$ crystals was not as shiny as e.g. that
of the $x = 0.85$ crystals. The higher $\Gamma_1(0)$ also results in
the lower amplitude and larger width of the Drude peak observed in
$\sigma_1$ of $x=0.9$ if compared to $x=0.85$ (inset of
Fig.~\ref{fig:Drudeweight}) which also results in a considerable
reduction of $I_{\sigma}$ of the correlated carriers seen in
Figs.~\ref{fig:Drudeweight} and \ref{fig:Neff} in the limit of low
$T$. Since the dynamical properties of the correlated carriers are
not influenced, the $x = 0.9$ sample still gives very valuable
information for our analysis.
\bibitem{Brinkman70}W.F. Brinkman and T.M. Rice, Phys. Rev. B {\bf 2}, 4302 (1970).
\bibitem{Kadowaki86} K. Kadowaki and S.B. Woods, Solid State Commun. {\bf 58}, 507 (1986).
\bibitem{Carter93}S.A. Carter, T.F. Rosenbaum, P. Metcalf, J.M. Honig, and J. Spalek, Phys. Rev. B {\bf 48}, 16841 (1993).
\bibitem{Mori99}T. Mori,  H. Mori, and S. Tanaka, Bull. Chem. Soc. Jpn. {\bf 72}, 179 (1999).
\bibitem{Watanabe91}Y. Watanabe, H. Seo, T. Sasaki, and N. Toyota, J.
Phys. Soc. Jpn.  {\bf 60}, 3608 (1991).




\end{thebibliography}
\end{document}